\documentclass{article}

\usepackage{arxiv}

\usepackage[utf8]{inputenc} 
\usepackage[T1]{fontenc}    
\usepackage{hyperref}       
\usepackage{url}            
\usepackage{booktabs}       
\usepackage{amsmath, amssymb, amsfonts}       
\usepackage{nicefrac}       
\usepackage{microtype}      
\usepackage{lipsum}		
\usepackage{graphicx}
\usepackage{natbib}
\usepackage{doi}

\title{An optimization framework for wind farm layout design using CFD-based Kriging model}

\author{ Zhenfan Wang \\
	School of Naval Architecture, Ocean and Civil Engineering\\
	Shanghai Jiao Tong University\\
	Shanghai 200240, China \\
 	\texttt{sjtu-wzf@sjtu.edu.cn} \\
	\And
         Yu Tu \\
	School of Naval Architecture, Ocean and Civil Engineering\\
	Shanghai Jiao Tong University\\
	Shanghai 200240, China \\
	\And
        \href{https://orcid.org/0000-0001-6097-7217}
	{\includegraphics[scale=0.06]{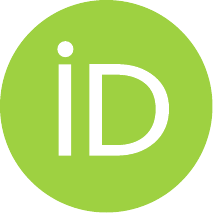}\hspace{1mm}Kai Zhang} \\
	School of Naval Architecture, Ocean and Civil Engineering\\
	Shanghai Jiao Tong University\\
	Shanghai 200240, China \\
	\texttt{kai.zhang@sjtu.edu.cn} \\
	\AND
	Zhaolong Han \\
	School of Naval Architecture, Ocean and Civil Engineering, Shanghai Jiao Tong University \\
	Shanghai 200240, China \\
	\And
 	Yong Cao \\
	School of Naval Architecture, Ocean and Civil Engineering, Shanghai Jiao Tong University \\
	Shanghai 200240, China \\
	\And
	Dai Zhou \\
	School of Naval Architecture, Ocean and Civil Engineering, Shanghai Jiao Tong University \\
	Shanghai 200240, China \\
}

\date{August 28,2023}

\renewcommand{\shorttitle}{Effects of yaw misalignment control on the aerodynamics of two tandem turbines}

\begin{document}
\maketitle

\begin{abstract}
Wind farm layout optimization (WFLO) seeks to alleviate the wake loss and maximize wind farm power output efficiency, and is a crucial process in the design and planning of wind energy projects.
Since the optimization algorithms typically require thousands of numerical evaluations of the wake effects, conventional WFLO studies are usually carried out with the low-fidelity analytical wake models, while the higher-fidelity computational-fluid-dynamics-based (CFD-based) methods are seldom used due to the excessive computational cost.
In this paper, we develop a self-adaptive optimization framework for wind farm layout design using CFD-based Kriging model to maximize the annual energy production (AEP) of wind farms. 
This surrogate-based optimization (SBO) framework uses latin hypercube sampling to generate a group of wind farm layout samples, based on which CFD simulations with the turbines modeled as actuator disks are carried out to obtain the corresponding AEPs.
This initial wind farm layout dataset is used to train the Kriging model, which is then integrated with an optimizer based on genetic algorithm (GA). 
As the optimization progresses, the intermediate optimal layout designs are again fed into the dataset to re-train the Kriging model.
Such adaptive update of wind farm layout dataset continues until the algorithm converges to the optimal layout design.
To evaluate the performance of the proposed SBO framework, we apply it to three wind farm cases under different wind distribution and terrains. 
Compared to the conventional staggered layout along the dominant wind direction, the optimized wind farm produces significantly higher total AEP, which is more evenly distributed among the turbines.
In particular, the SBO framework requires significantly smaller number of CFD calls to yield the optimal layouts that generates almost the same AEP with the direct CFD-GA method.
Further analysis on the velocity fields show that the optimization framework always attempts to locate the downstream turbines away from the the wakes of upstream ones along the dominant wind directions.
The proposed CFD-based surrogate model provides a more accurate and flexible alternative to the conventional analytical-wake-model-based methods in WFLO tasks, and has the potential to be used for designing efficient wind farm projects.
\end{abstract}


\section{Introduction}
\label{sec:intro}

An increasing number of countries have committed to developing sustainable energy as substitutes for fossil fuels due to pressing environmental issues. 
As wind energy is one of the most economically competitive and abundant renewable energy, wind farms have been installed worldwide to harvest this clean energy resources. However, the interaction between turbines within wind farms, particularly the wake effects generated by upstream turbines, can have a substantial impact on the performance of downstream turbines. These wake effects lead to lower wind speeds and increased turbulence intensity, resulting in reduced power generation. It is reported that the average power loss reaches $10\% - 20\%$ in some large offshore wind farms due to the the wake effects \citep{barthelmie2009modelling}. 

As an effective means to alleviate the wake effects within wind farms at the initial designing phase, wind farm layout optimization (WFLO) has been a subject of extensive researches \citep{reddy2020wind,dong2021intelligent,thomas2022comparison}.
Table \ref{tab:wflo} presents a selective summary of the existing studies on WFLO in terms of the optimization algorithm and the wake model.
Both the gradient-based methods such as sparse nonlinear optimizer (SNOPT) and Sequential Convex Programming (SCP), and the heuristic search techniques such as the greedy algorithm, genetic algorithm (GA), particle swarm optimization algorithm (PSO), ant colony optimization (ACO), have been applied in these studies.
Two main types of wake models are used in WFLO, namely the analytical wake models and the numerical wake models. 

\begin{table}
\centering
    \caption{\label{tab:wflo}A review of the wind farm layout optimization studies}
    \begin{tabular}{c c c}
         \toprule
           Reference & Optimization algorithms & Wake models  \\
         \midrule
           \citet{mosetti1994optimization}  & GA & Jensen \\
           \citet{ozturk2004heuristic} & Greedy & - \\
           \citet{grady2005placement} & GA & Jensen \\  
           \citet{mora2007evolutive} & GA & Jensen \\ 
           \citet{zhang2011fast} & Greedy & Jensen \\
           \citet{erouglu2012design} & ACO & Jensen\\
           \citet{chen2013wind} & GA & Frandsen \\
           \citet{park2015layout} & SCP & Jensen \\
           \citet{shakoor2015modelling} & GA & Jensen \\
           \citet{hou2016optimization} & PSO & Jensen \\
           \citet{gebraad2017maximization} & SNOPT & Floris wake model \\
           \citet{parada2017wind} & GA & BP \\
           \citet{pillai2017application} & GA \& PSO & Larsen \\
           \citet{kirchner2018realistic} & GA & Gaussian\\
           \citet{stanley2019massive} &SNOPT & Gaussian \\
           \citet{quan2019greedy} & Greedy & Jensen\\
           \citet{cruz2020wind} & GA & CFD-ADM \\
           \citet{antonini2020optimal} & SQP & CFD-ADM \\
           \citet{stanley2020wind} & SNOPT & BP \\       
           \citet{gagakuma2021reducing} & SNOPT & Floris wake model \\
           \citet{liu2021genetic} & GA & Ishihara \\
           \citet{thomas2022wake} & SNOPT \& ALPSO & Gaussian \\
         \bottomrule
    \end{tabular}
\end{table} 

Analytical wake models such as the Jensen wake model \citep{jensen1983note}, the Gaussian model \citep{larsen1988simple}, the Frandsen wake model \citep{frandsen2006analytical} are most widely applied in WFLO problems due to the high computational efficiency. 
These models employ simplified analytical functions based on momentum conservation or empirical relations to describe the turbine's wake characteristics. 
The most used Jensen model assumes that the wake expands linearly, and the velocity deficit is determined by the distance behind the turbine, wind speed and induction parameter.
The Gaussian wake model assumes that the velocity deficit follows a Gaussian distribution. It provides a more realistic representation of the wake's shape compared to the linear wake expansion in Jensen's model.
The Frandsen wake model incorporates both linear and Gaussian wake expansions, and offers more accurate predictions by accounting for the initial wake width.
Although these analytical models provide valuable insights into the wake development behind wind turbines, they are simplifications of the complex physics involved in wake interactions, and can be less accurate in assessing the wake effects.

Numerical simulations based on computational fluid dynamics (CFD) presents a higher-fidelity method for understanding the aerodynamics of wind turbines.
Depending on the way the turbines are modeled, the CFD-based methods can be classified as blade-resolved simulation, actuator line model, and actuator disk model.
Blade-resolved simulations provide detailed insights into the flow characteristics around individual blades, including the effects of turbulence, separation, and dynamic stall \citep{mittal2016blade,liu2017establishing,de2022blade,zhang2023comparative}. However, the huge computational resources required for resolving the blade boundary layer preclude its use for large-scale wind farm studies.
The actuator line model (ALM) treats the rotor blades as lines distributed with actuator points, on which the sectional drag and lift forces are projected. 
Since these aerodynamic forces are calculated based on tabulated airfoil data, this approach removes the computational complexity associated with resolving the rotor blades. 
ALM has been regarded as the state-of-the-art tool for wind farm simulations \citep{stevens2017flow,stevens2018comparison,shapiro2022turbulence}. 
Nevertheless, in the case of WFLO where iterative evaluations of the velocity deficit are required, ALM is still prohibitively expensive.

Another popular method in the actuator-type modeling technique is the actuator disk model (ADM).
This model further simplifies the rotor swept area as permeable disk.
The thrust and torque of the rotor are added to the actuator zone as source terms, which generate pressure drops and velocity plunges across the disk area in the CFD simulations \citep{sanderse2011review,svenning2010implementation}.
Compared with ALM, the more intricate flow structures such as root and tip vortices can not be captured by ADM.
Nevertheless, the ADM has been shown to depict wake effects with sufficient accuracy \citep{martinez2015large}.
In recent years, CFD simulations with ADM has been applied in WFLO problems.
\citet{antonini2020optimal} presented a gradient-based algorithm with adjoint method for gradient calculation to solve the optimization of wind farm layout using ADM. 
The authors claimed that this CFD-based method is able to accurately simulate wake effects and terrain-induced flow characteristics. 
\citet{cruz2020wind} applied the heuristic genetic algorithm to optimize the layout of wind farm with terrain effects using CFD.
While the authors also commended the superior accuracy of ADM over the analytical wake model in WFLO tasks, they noted that significant amount of computational resources are required for the optimization algorithm to converge to the optimal layout design.

To alleviate the computational challenges of using high-fidelity CFD-based methods in WFLO problems, we propose to apply the Kriging model as a surrogate to accelerate the optimization process. 
The Kriging model, also known as Gaussian Progress Regression (GPR), was initially developed as a geostatistical method for spatial interpolation and prediction of correlated data \citep{krige1951statistical}.
It estimates the unknown values at unobserved locations by taking into account the observed values at nearby locations and their spatial relationships. 
Once this surrogate model is trained, the wake effects and the energy production of a particular layout can be obtained at a low computational cost and without the need of new CFD simulations. 
The applicability of Kriging model has been demonstrated in various disciplines such as image segmentation \citep{karl2010spatial}, hull shape optimization \citep{casella2020surrogate}, aero-elastic tailoring of bridge decks \citep{montoya2022aero}, just to name a few.
However, the feasibility of using Kriging model for high-fidelity wind farm design has not been reported before.

In this paper, we develop a self-adaptive optimization framework for  high-fidelity wind farm layout design using surrogate-based optimization (SBO) method. 
Using this framework, we optimize the siting of each turbine within a restricted area to maximize wind farm's annual energy production (AEP) with reduced computational cost. 
The rest of the paper is organized as follows: Section \ref{sec:framework} outlines the formulation of the WFLO problems and the self-adaptive WFLO framework, including the wind farm layout dataset, the numerical method, and the surrogate-based optimization (SBO) method using Kriging. 
In Section \ref{sec:cases}, we present three wind farm cases under different wind distributions and terrains to demonstrate the performance and advantages of our proposed framework. 
Finally, Section \ref{sec:conclusion} presents the conclusions and potential future research directions.

\section{Self-adaptive WFLO framework}
\label{sec:framework}
\begin{figure*}
    \centering
    \includegraphics[width=1.0\textwidth]{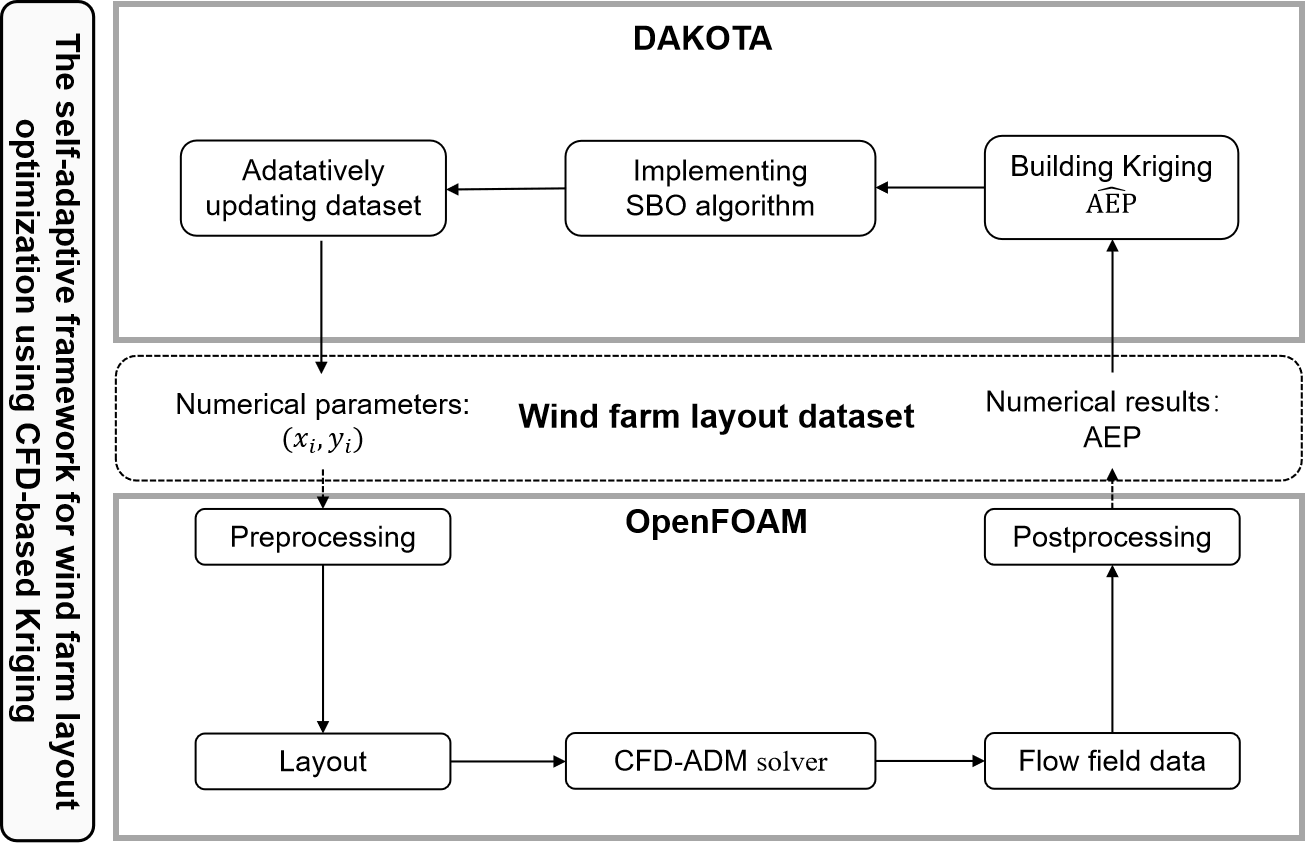}
    \caption{Flow chart of the self-adaptive WFLO framework. }
    \label{fig:framework}
\end{figure*}

In this section, we present the details about the proposed wind farm layout optimization (WFLO) framework.
The formulation of the WFLO problem is formally introduced in \S \ref{sec:problem}.
Next, we outline the three main procedures involved in the framework, i.e., the adaptive wind farm layout dataset (\S \ref{sec:dataset}), the prediction of AEP using CFD simulations (\S \ref{sec:cfd}), and the construction of Kriging model for surrogate-based optimization (\S \ref{sec:sbo}).
Figure \ref{fig:framework} illustrates the self-adaptive WFLO framework. 
The open-source CFD toolbox OpenFOAM is used for wind farm CFD simulations and AEP prediction. 
The construction of Kriging model and implementation of SBO are carried out using optimization toolbox DAKOTA. Python and shell glue scripts are created to couple these toolboxes.    

\subsection{Formulation of WFLO problem}
\label{sec:problem}
The objective of the WFLO problem is to find the optimal location of each wind turbine to maximize annual energy production (AEP) of the wind farm:
$$ \mathrm{AEP}=\frac{365\times3+366}{4}\times24\sum_{i=1}^{n}\sum_{j=1}^{m}P_{j}(x_{i},y_{i})f_{j},$$  
where $n,m$ denote numbers of turbine and wind directions respectively, $P_{j}$ is turbine's power output under wind direction $j$ , $f_{j}$ is the frequency of wind direction $j$. $(x_{i},y_{i})$ is the coordinate of turbine $i$ , and is the design variable in this study. 
To reduce computational cost, we choose to solve a discrete optimization problem by splitting the siting area into square cells as candidate positions for wind turbines \citep{bai2022wind}. 
It was shown in \citet{thomas2022comparison} that the limitation on potential turbine locations does not necessarily preclude such methods from finding optimal layouts.
In WFLO problem, additional inter-distance requirement is defined as:
$${\rm subject \ to}\ d_i \geq d_{min},$$
where $d_i$ is the inter-distance between turbines, and must be greater than or equal to $d_{min} = 2D$.

\subsection{Adaptive wind farm layout dataset}
\label{sec:dataset}
The wind farm layout parameters and the corresponding AEP acquired from CFD simulations form the wind farm layout dataset. It is comprised of two groups of data, the initial sampling data and iteratively updated data. The distribution of initial samples determines the performance of the framework. Therefore, it needs to be equally distributed throughout design space. We use latin hypercube sampling (LHS) method \citep{mckay2000comparison} to produce the initial samples of layout parameters, $\boldsymbol{X} = \{\boldsymbol{x}^{(1)},\boldsymbol{x}^{(2)},...,\boldsymbol{x}^{(n)}\}^T$. The $\boldsymbol{x}^{(i)} = \{(x^{(i)}_{1},y^{(i)}_{1}), (x^{(i)}_{2},y^{(i)}_{2}), ..., (x^{(i)}_{k},y^{(i)}_{k})\}$ represents turbines' horizontal coordinate set in the $i$th sample of layout.  
CFD simulations are carried out based on the layout parameters, and the annual energy production (AEP) is collected in $\boldsymbol{Y} = \{\mathrm{AEP}^{(1)}, \mathrm{AEP}^{(2)}, ..., \mathrm{AEP}^{(n)}\}^T$. 
The iterative update of dataset is explained in section \ref{sec:sbo}.

\subsection{Wind farm CFD simulations}
\label{sec:cfd}
In this study, CFD simulations are carried out to model the wake interactions with the rotors treated as actuator disks. 
The power production of each turbine is predicted using the the ADM-RANS simulations, as illustrated below.

\subsubsection{Governing equations}
\label{sec:gov}
We numerically solve for the flows over the wind turbines by employing Reynolds-averaged Navier-Stokes (RANS) formulation 
\begin{subequations}
    \label{equ:rans}
    \begin{align}
          \frac{\partial \overline{u}_i}{\partial x_i} & = 0, \\
         \rho \frac{\partial \overline{u}_i}{\partial t} + \rho\frac{\partial(\overline{u}_i\overline{u}_j)}{\partial x_j} & = -\frac{\partial \overline{p}}{\partial x_i}+\frac{\partial}{\partial x_j}\left(\mu\left(\frac{\partial \overline{u}_i}{\partial x_j}+\frac{\partial \overline{u}_j}{\partial x_i}\right)-\rho\overline{u^{'}_{j} u^{'}_{i}}\right) + f_i,
    \end{align}
\end{subequations}
where $x_{i}$ is Cartesian space coordinate, $\overline{u}_{i}$ and $\overline{p}$ are the temporally averaged velocity and pressure, respectively.
$\rho$ and $\mu$ are the air density and dynamic viscosity, $f_i$ represents the source term from the actuator disk.
The Reynolds stress emerged from the time-average process $\rho\overline{u_{j}^{'}u_{i}^{'}}$ is modeled using the $k-\epsilon$ turbulence model with the introduction of two new variables, i.e., the turbulence kinematic energy $k$, and the dissipation rate $\epsilon$.
The standard transport equations for the two new variables read as
\begin{subequations}
    \label{equ:turbulence}
    \begin{align}
        \rho\frac{\partial k}{\partial t}+\rho\frac{\partial (\overline{u}_{i}k)}{\partial x_i} &=\frac{\partial}{\partial x_j}\left(\frac{\mu_t}{\sigma_k}\frac{\partial k}{\partial x_j}\right)+P_k-\rho\epsilon , \\
        \rho\frac{\partial\epsilon}{\partial t}+\rho\frac{\partial(\overline{u}_{i}\epsilon)}{\partial x_i} &=\frac{\partial}{\partial x_j}\left(\frac{\mu_t}{\sigma_\epsilon}\frac{\partial\epsilon}{\partial x_j}\right)+C_{1\epsilon}\frac{\epsilon}{k}P_k - C_{2\epsilon}\frac{\epsilon^2}{k}\rho ,
    \end{align}
\end{subequations}
with 
\begin{subequations}
\label{equ:pkmu}
    \begin{align}
        P_k &= -\rho\overline{u_{j}^{'}u_{i}^{'}}\frac{\partial u_j}{\partial x_i} , \\
        \mu_t &= -\rho C_\mu \frac{k^2}{\epsilon},
    \end{align}
\end{subequations}
where $C_\mu$, $\sigma_k$,  $\sigma_\epsilon$, $C_{1\epsilon}$, $C_{2\epsilon}$ are the five constants in the $k-\epsilon$ turbulence model and are set as default values in OpenFOAM.

The RANS equations outlined above are discretized using the finite volume method with second-order numerical schemes in both space and time. 
A modified version of the simpleFoam solver (OpenFOAM package) is used for solving for the steady-state flow over the wind turbines.
The modification involves adding the ADM source term into the RANS equations, as we discuss in detail next.

\subsubsection{Actuator disk model}
\label{sec:adm}
The actuator disk model (ADM) simulates the turbine rotor as a thin, permeable disk, and the momentum transferred from rotor to the flow is added to the flow as volumetric source term $f_i$, as mentioned in equation \ref{equ:rans}.
To implement wind farm CFD simulations, thrust and torque of each rotor are required. 
Ideally, these aerodynamic quantities can be obtained by looking up the thrust and torque curves provided by wind turbine manufacturers, given the theoretical inflow velocity.
However, in the presence of the wake effects within wind farm, the theoretical inflow velocity for the downstream turbine is not known \emph{a prior}.

\begin{figure}
    \centering
    \includegraphics[width=1.0 \textwidth]{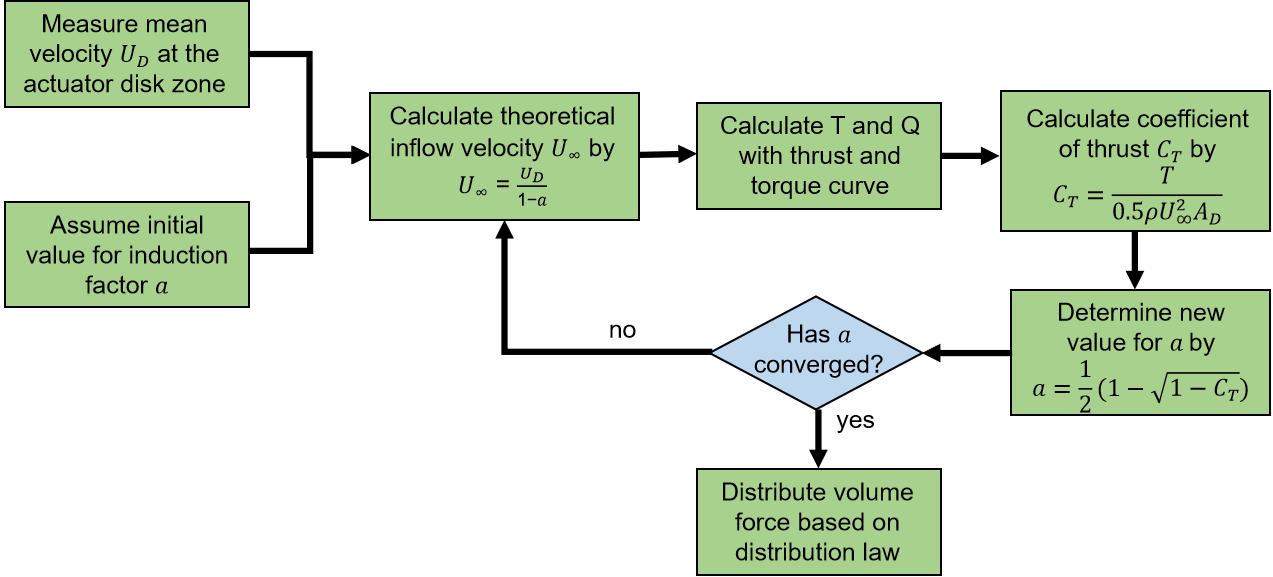}
    \caption{Workflow to calculate inflow velocity, thrust and torque in the ADM code}
    \label{fig:inflowvelocity}
\end{figure}

To circumvent this problem, we follow the approach in \citet{richmond2019evaluation}, where the thrust and torque curves are inserted into the flow solver, and the theoretical inflow velocity is calculated in an iterative manner.
This method, which is summarized in figure \ref{fig:inflowvelocity}, is based on the one-dimensional actuator disk theory \citep{burton2011wind}:
\begin{subequations}
\label{equ:admtheory}
    \begin{align}
        & U_{D} = U_{\infty}(1-a), \\
        & C_{T} = \frac{T}{\frac{1}{2}\rho U^{2}_{\infty}A_{D}}, \\
        & C_{T} = 4a(1-a), \\
        & a = \frac{1}{2}(1-\sqrt{1-C_T}),
    \end{align}
\end{subequations}
where $U_D$ is the velocity averaged over the disk region, $C_T$ is the coefficient of thrust, $A_D$ is the area of actuator disk, and $a$ is axial induction factor.
This equation is solved iteratively within the modified simpleFoam code, for each turbine at each solver iteration, until the convergence criteria is met.

After the thrust and torque of each rotor are obtained, the volume forces are distributed along the radial direction following the Goldstein optimum 
 \citep{goldstein1929vortex}:
\begin{subequations}
\label{equ:distribution}
    \begin{align}
        f_{ix} &= A_{x}r^*\sqrt{1-r^*}, \\
        f_{i\theta} &= A_\theta \frac{r^*\sqrt{1-r^*}}{r^*(1-r^{'}_{h})+r^{'}_{h}},
    \end{align}
\end{subequations}
with
\begin{subequations}
\label{equ:distributionpara}
    \begin{align}
        r^* &= \frac{r^{'}-r^{'}_h}{1-r^{'}_h}, \ \ r^{'} = \frac{r}{R_P}, \\
        A_x &= \frac{105}{8}\frac{T}{\pi\Delta(3R_H+4R_P)(R_P-R_H)}, \\
        A_\theta &= \frac{105}{8}\frac{Q}{\pi\Delta R_P(3R_P+4R_H)(R_P-R_H)},
    \end{align}
\end{subequations}
where $f_{ix}$ is axial force, $f_{i\theta}$ is tangential force, $r$ is the distance between the point and disk center, $R_P$ is the external radius of disk, $R_H$ is the internal radius of disk, $T$ is rotor's thrust, $Q$ is rotor's torque, and $\Delta$ is the thickness of disk.

Finally, the objective function, $\mathrm{AEP}$, is predicted by the sum of power of each individual cell within actuator disk zones as 
\begin{equation}
\label{equ:power}
    P = \sum_{i=1}^{k}P_{i} = \sum_{i=1}^{k}F_{i}U_{i}
\end{equation}
where $F_i$ and $U_i$ are vectors of volume force and velocity in each individual cell, and $k$ is total number of cells in the disk zone.

\subsubsection{Computational setup}
\label{sec:setup}
\begin{figure}
    \centering
    \includegraphics[width=1.0 \textwidth]{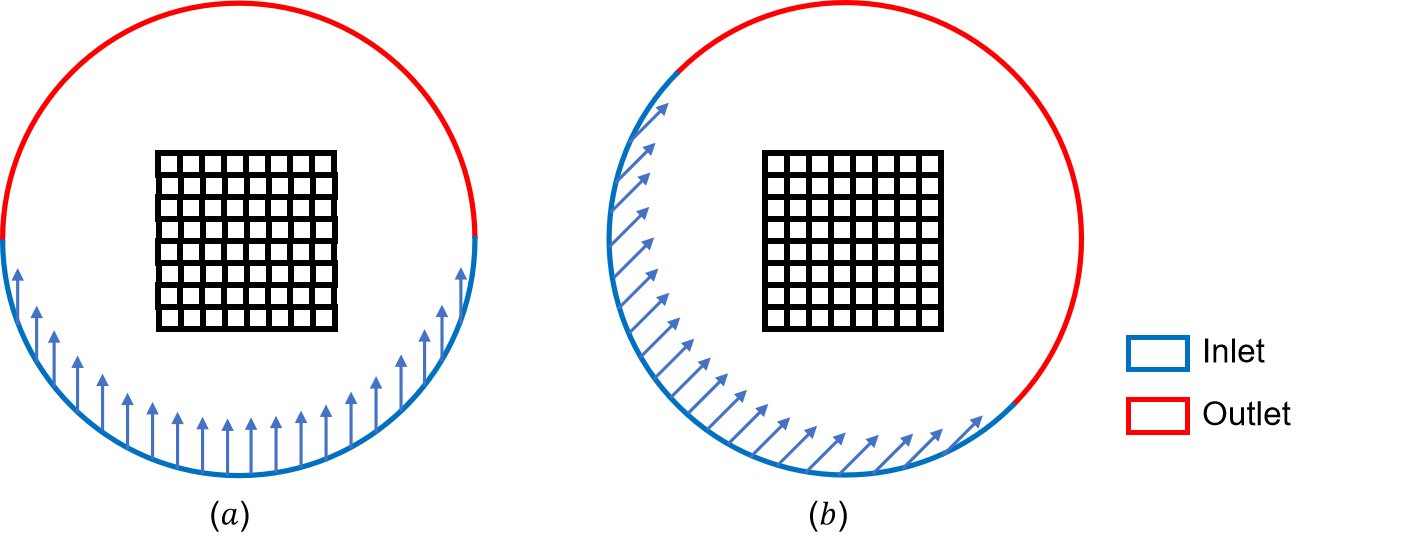}
    \caption{Top views of the circular computational domain for a wind farm. $(a)$ assigning inlet or outlet boundary for wind direction of $180^\circ$, $(b)$ assigning inlet or outlet boundary for wind direction of $225^\circ$.}
    \label{fig:inletoutlet}
\end{figure}

We adopt a circular computational domain that allows for simulations with different inflow directions, as illustrated in figure \ref{fig:inletoutlet}.
The wind turbines are restricted within a rectangular subdomain in the center.
The diameter of the circular computational domain is set as 3 times the length of the center rectangle.
In the vertical direction, the height of the computational domain is set as $8D$, where $D$ is the diameter of the rotor.
The inlet velocity is prescribed as $U_{in}=U_{\infty}(z/z_0)^{\alpha}$, where $z_0$ is the height of the turbine hub, $U_{\infty} = 11.4$ m/s is the reference velocity at the turbine hub, and $\alpha=0.14$ is the shear rate of the velocity profile.  
At the outlet, a reference pressure $p_{\infty}=0$ is set, with zero gradient condition for the velocity.
The bottom side of the computational domain is treated as wall, and the top boundary as slip condition.

The mesh of the computational domain is created using OpenFOAM's native blockMesh tool.
To guarantee that ADM’s source terms are loaded into the flow, meshes near actuator disk zone are refined. The element size in disk region is $6$ m ($0.05D$).
For a simulation with 8 turbines, the total number of control volumes is around $2 \times 10^5 -6\times 10^5$. 

\subsection{Surrogate-based optimization method} 
\label{sec:sbo}
\subsubsection{Kriging model}
\label{sec:kg}
The Kriging model \citep{krige1951statistical} is employed in this framework to build the surrogate model. It predicts the response values at unknown points based on a set of sample input data $\boldsymbol{X} = \{ \boldsymbol{x}^{(1)},\boldsymbol{x}^{(2)},...,\boldsymbol{x}^{(n)}\}^{T}$, and the observed values $\boldsymbol{Y} = \{ y^{(1)},y^{(2)},...,y^{(n)}\}^{T}$.

The Kriging emulator, $\hat{f}(x)$, is defined in the following equation \citep{dalbey2022dakota,adams2020dakota}:
\begin{equation}
    \hat{f}(\boldsymbol{x}) = g(\boldsymbol{x})^{T}\beta + \epsilon(\boldsymbol{x}),
\end{equation}
where $\hat{\ }$ denotes the Kriging approximation, $\boldsymbol{x}$ is the vector with all the design variables, $g(\boldsymbol{x})\beta$ is the vector of trend functions, $\epsilon(\boldsymbol{x})$ is Gaussian process error model which has zero uncertainty at the training points. 
The covariance matrix $\epsilon (\boldsymbol{x})$ is modeled using the correlation matrix as
\begin{equation}
    \mathrm{Cov}[\epsilon(\boldsymbol{x}^{(i)}),\epsilon(\boldsymbol{x}^{(j)})] = \sigma^{2}r(\boldsymbol{x}^{(i)},\boldsymbol{x}^{(j)}),
\end{equation}
where $r(\boldsymbol{x}^{(i)},\boldsymbol{x}^{(j)})$ is the correlation functions between samples $\boldsymbol{x}^{(i)}$ and $\boldsymbol{x}^{(j)}$. 

In this framework, the Gaussian correlation function is used to calculate the correlation vector and matrix. The correlation vector and matrix are constructed by the following equations:
\begin{subequations}
\begin{align}
    & r(\boldsymbol{x}^{(i)},\boldsymbol{x}^{(j)}) = \mathrm{exp}(-\sum_{k=1}^m\theta_k|{x}^{(i)}_{k}-{x}^{(j)}_{k}|^{2}), \ \ \ i,j\ =\ 1,2,...,n \\
    & \boldsymbol{R} = \begin{pmatrix}
        r(\boldsymbol{x}^{(1)},\boldsymbol{x}^{(1)}) & \cdots & r(\boldsymbol{x}^{(1)},\boldsymbol{x}^{(n)}) \\
        \vdots & \ddots & \vdots \\
        r(\boldsymbol{x}^{(n)},\boldsymbol{x}^{(1)}) & \cdots & r(\boldsymbol{x}^{(n)},\boldsymbol{x}^{(n)})
    \end{pmatrix},
\end{align}
\end{subequations}
where $m$ is the number of dimensions in the search space; $n$ is number of samples; $\theta$ is hyperparameter that affects correlations between samples. 

The hyperparameters $\theta$ is estimated via Maximum Likelihood Estimation (MLE) method. To simplify MLE, the natural logarithm is used and constant terms are omitted:
\begin{equation}
\label{equ:mle}
    \mathrm{ln}(L) = -\frac{n}{2}ln|\boldsymbol{R}|-\frac{1}{2}ln(\hat{\sigma}^2),
\end{equation}
where $\hat{\sigma}^2=(\boldsymbol{Y-G}\beta)^T\boldsymbol{R}^{-1}(\boldsymbol{Y-G}\beta)/n$ is the MLE of $\sigma^2$ and $\boldsymbol{G}$ is an $n\times q$ matrix with rows $\boldsymbol{g(x)}_i^T$ (the trend function evaluated at point $i$).  Maximizing the equation (\ref{equ:mle}), we obtain the MLE of hyperparameter $\theta$ which in turn determines $\hat{\sigma^2}$.

When MLE is done, we obtain the Kriging emulator of unknown truth function. The predicted value $\hat{f}(\textbf{x})$ at a new point $\textbf{x}$ is:
\begin{equation}
    \hat{f}(\boldsymbol{x}) = g(\boldsymbol{x})^{T}\beta + \boldsymbol{r(x)}^T\boldsymbol{R}^{-1}(\boldsymbol{Y-G}\beta),
\end{equation}
where $\boldsymbol{r(x)}$ is the vector of correlation between $\boldsymbol{x}$ and each of training points.

Based on the wind farm layout dataset, a high-fidelity Kriging model $\widehat{\mathrm{AEP}}(x_1,y_1,x_2,y_2,...,x_n,y_n)$ for wind farm AEP prediction is built with the aforementioned process and coupled to optimization algorithms. The adaptive SBO method is introduced in detail in section \ref{sec:algorithms}.

\subsubsection{Adaptive surrogate-based optimization}
\label{sec:algorithms}
The surrogate-based optimization algorithm requires to construct one surrogate over the whole design space. The global surrogate is then coupled to optimization algorithms. Surrogate model built with initial sampling dataset is usually not accurate enough to be applied in optimization problem. For this reason, various methods to update truth dataset have been explored. An effective method is to infill the optimal design of $\hat{f}$ during last iteration with its truth response dataset. 
Given its advantage of low additional computational cost \citep{thelen2016surrogate}, this infilling strategy is employed in this study. The main steps of the adaptive SBO algorithm are as follows:

\begin{enumerate}
    \item  Generate a group of initial sampling parameters using LHS method.
    \item  Apply the simulation models to obtain the truth response for all sampling parameters.
    \item  Choose proper approximation method (Gaussian Process, also known as Kriging model in this framework) to construct the surrogate model based on the sampling parameters with their truth response.   
    \item  Couple surrogate model to the optimization algorithm and obtain optimum of the surrogate.
    \item  Check if maximum number of iteration is exceeded or optimal design converges. If not, repeat (2) - (4).
\end{enumerate}

\begin{figure}
    \centering
    \includegraphics[width=1.0\textwidth]{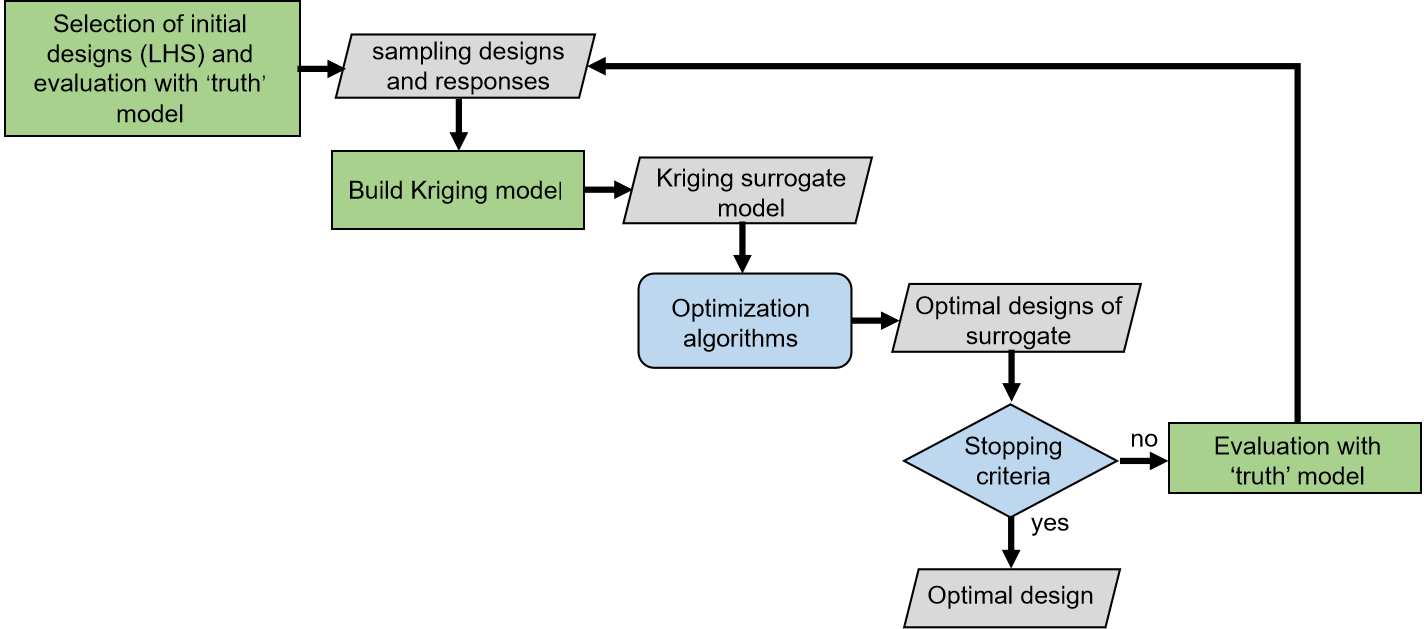}
    \caption{Workflow of the adaptive SBO method}
    \label{fig:workflow}
\end{figure}
The workflow of this adaptive SBO method is illustrated in figure \ref{fig:workflow}. As the optimization evolves, the adaptive samples with their truth response are infilled into the dataset and surrogate becomes increasingly accurate. Finally, the optimal solution is obtained.

Genetic algorithm (GA) is selected as the optimization algorithm in this study due to its wide application in WFLO problems as shown in table \ref{tab:wflo}. 
The process of GA includes population initialization, selection of the most fitted individuals, application of some natural processes such as crossover and mutation, and population reproduction \citep{liu2021genetic}. The detailed information of these steps is shown as follows:
\begin{enumerate}
    \item Population initialization.  Initialize a random population and then compute the value of the fitness function for each individual.
    \item Selection. Individuals in population which do not satisfy constraints will be eliminated or multiplied by a penalty parameter. Then the algorithm selects suitable parents from individuals based on their fitness value. These parent individuals are further used for mating and generating new population. 
    \item Crossover and mutation. Randomly select two parent individuals. Then the process of crossover and mutation are performed according to crossover and mutation rate respectively. This step aims at creating offspring individuals and preventing GA from converging to local optimum solution.  
    \item Check convergence. If the convergence criteria is satisfied or maximum number of evaluations is exceeded, the iteration ends. Otherwise, repeat step II.
\end{enumerate}

\section{Case studies}
\label{sec:cases}
In this section, three wind farm cases with different wind directions and terrains are studied using the self-adaptive optimization framework. 
As a reference, the more computationally expensive CFD-GA optimization (without surrogate model) is also carried out.
The optimized results are then discussed and compared to the conventional staggered layout along the major wind direction (baseline) to assess the performance of the developed framework. The self-adaptive optimization framework settings for the three cases are summarized in table \ref{tab:framesettings}. Detailed case descriptions are presented in \S \ref{sec:description}. Results and discussions are presented in \S \ref{sec:results}.
\begin{table}
\centering
\caption{\label{tab:framesettings}Settings for the self-adaptive framework.}
\begin{tabular}{ c c }
         \toprule
          Parameters & Value   \\
         \midrule
          Initial dataset size  & 360   \\
          Population size & 50   \\
          Crossover rate & 0.95   \\
          Mutation rate & 0.15  \\
          Maximum iterations & 1000  \\
         \bottomrule
\end{tabular}
\end{table} 

\subsection{Case descriptions}
\label{sec:description}
\begin{figure}
    \centering
    \includegraphics[width=0.5\textwidth]{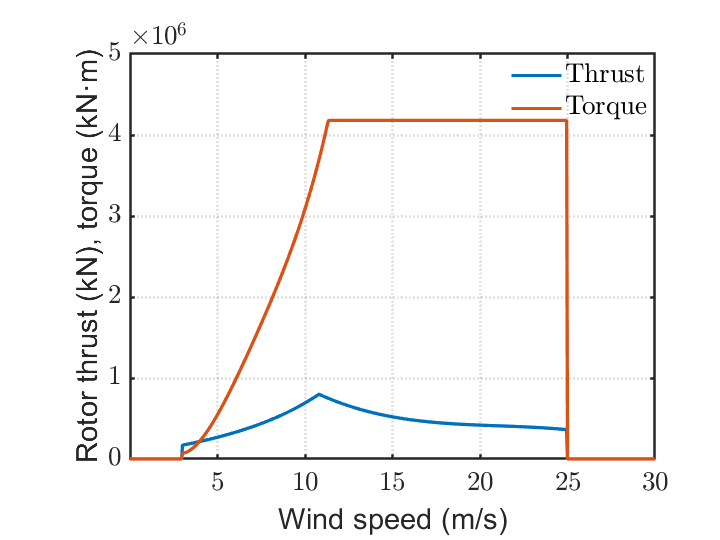}
    \caption{Thrust and torque curves for NREL 5 MW wind turbine }
    \label{fig:nrel5mw}
\end{figure}

We use the NREL 5 MW reference turbine \citep{jonkman2009definition} as the wind turbine model in this study.
The key parameters in the numerical modeling are listed in the table \ref{tab:nrel}. 
The thrust and torque curves of the NREL 5 MW wind turbine as shown in figure \ref{fig:nrel5mw} are inserted into the ADM code as mentioned in figure \ref{fig:workflow}.

\begin{figure}
    \centering
    \includegraphics[width=0.5\textwidth]{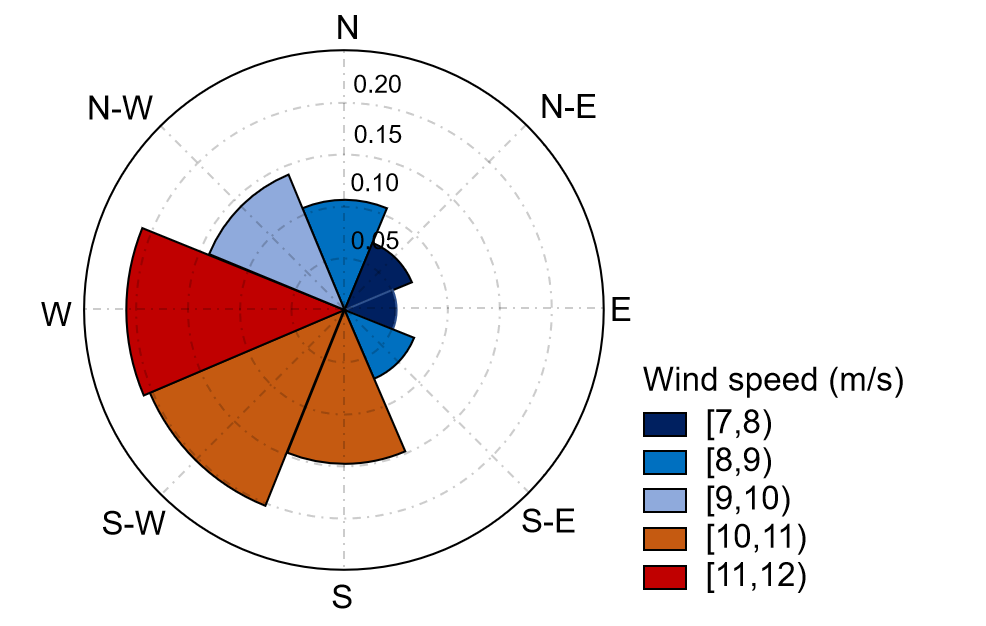}
    \caption{Wind rose for case II and case III}
    \label{fig:windrose}
\end{figure}

We optimize the layout of an 8-turbine wind farm within an $8D \times 8D$ rectangle, which is decomposed into $8 \times 8$ cells. The 81 nodes are prescribed as the potential positions for the eight turbines. 
We evaluate the performance of the proposed WFLO framework in three representative cases as shown in table \ref{tab:cases}.
In Case I, only one wind direction with $U_{\infty}=$ 11.4 m/s is considered.
In Case II, a wind rose with eight wind directions (shown in figure \ref{fig:windrose}) is taken into consideration.
On this basis, case III further considers the effects of the terrain on the wind farm layout by the addition of an Gaussian-shaped hill with the height of 150 m on the bottom boundary.

\begin{table}
\centering
    \caption{\label{tab:nrel}Key parameters of NREL 5 MW wind turbine}
    \begin{tabular}{c c}
         \toprule
          Parameters & Value \\
         \midrule
          Diameter (D)  & 126 m   \\
          Hub height & 90 m   \\
          Rated wind speed & 11.4 m/s   \\
          Rated power & 5 MW  \\
          Cut-in speed & 3 m/s  \\
          Cut-out speed & 25 m/s  \\
         \bottomrule
    \end{tabular}
\end{table}

\begin{table}
\centering
    \caption{\label{tab:cases}Definition for 3 WFLO cases}
    \begin{tabular}{c c c c}
         \toprule
                    & Case I   & Case II  & Case III  \\
         \midrule
           Turbine numbers & 8 & 8 & 8  \\
           Restricted area & $8D \times 8D$ & $8D \times 8D$ & $8D \times 8D$ \\
           Wind directions & 1 & 8 & 8  \\
           Terrains & flat terrain & flat terrain & Gaussian-shaped hill \\
         \bottomrule
    \end{tabular}
    \label{tab:cases}
\end{table}

\subsection{Optimization results and discussion}
\label{sec:results}

\begin{figure}
    \centering
    \includegraphics[width=1.0\textwidth]{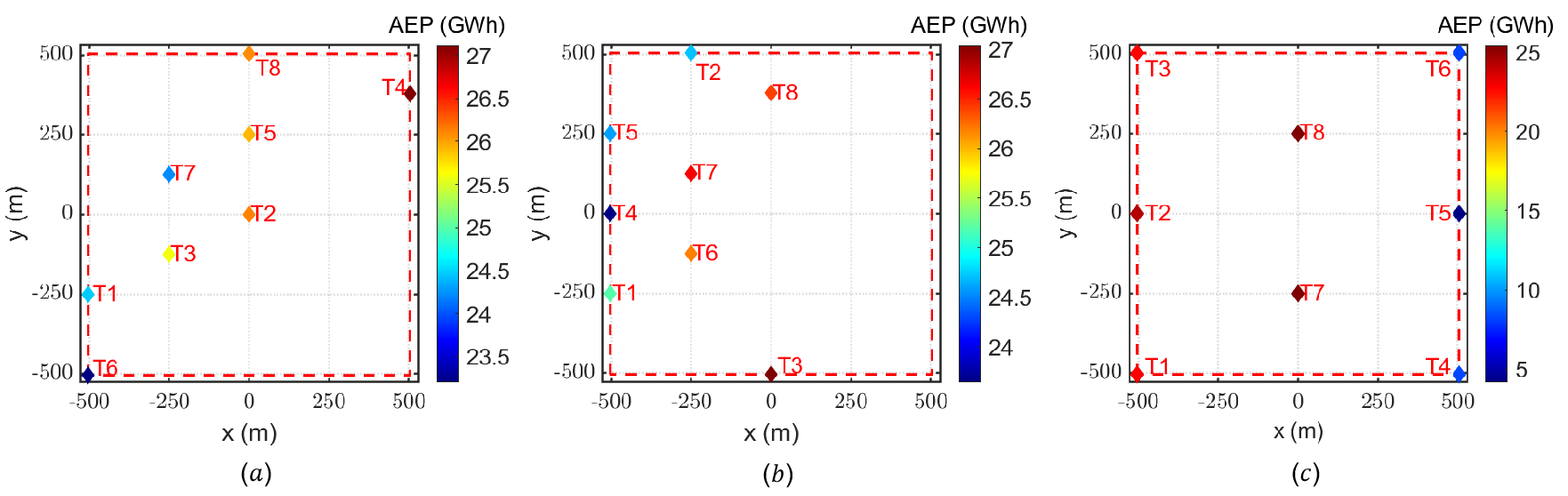}
    \caption{Case I: AEP distribution for layouts obtained from $(a)$ SBO $(b)$ CFD-GA and $(c)$ baseline. The turbines are colored based on their individual AEPs.}
    \label{fig:case1layout}
\end{figure}

\begin{figure}
    \centering
    \includegraphics[width=1.0\textwidth]{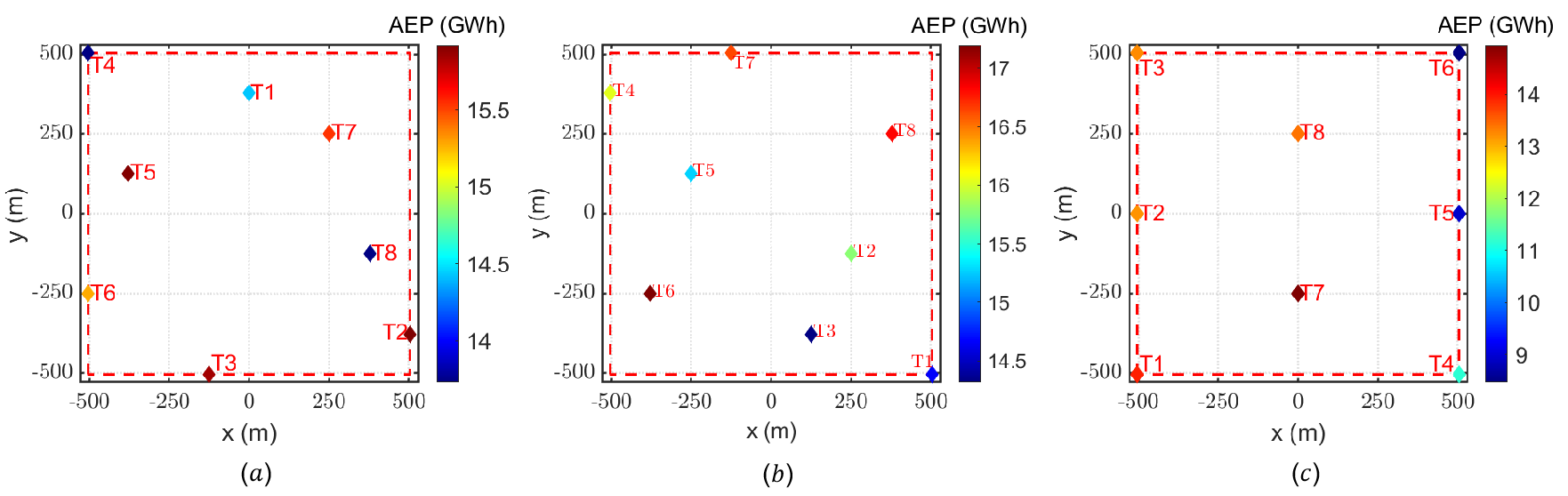}
    \caption{Case II: AEP distribution for layouts obtained from $(a)$ SBO $(b)$ CFD-GA and $(c)$ baseline.}
    \label{fig:case2layout}
\end{figure}

\begin{figure}
    \centering
    \includegraphics[width=1.0\textwidth]{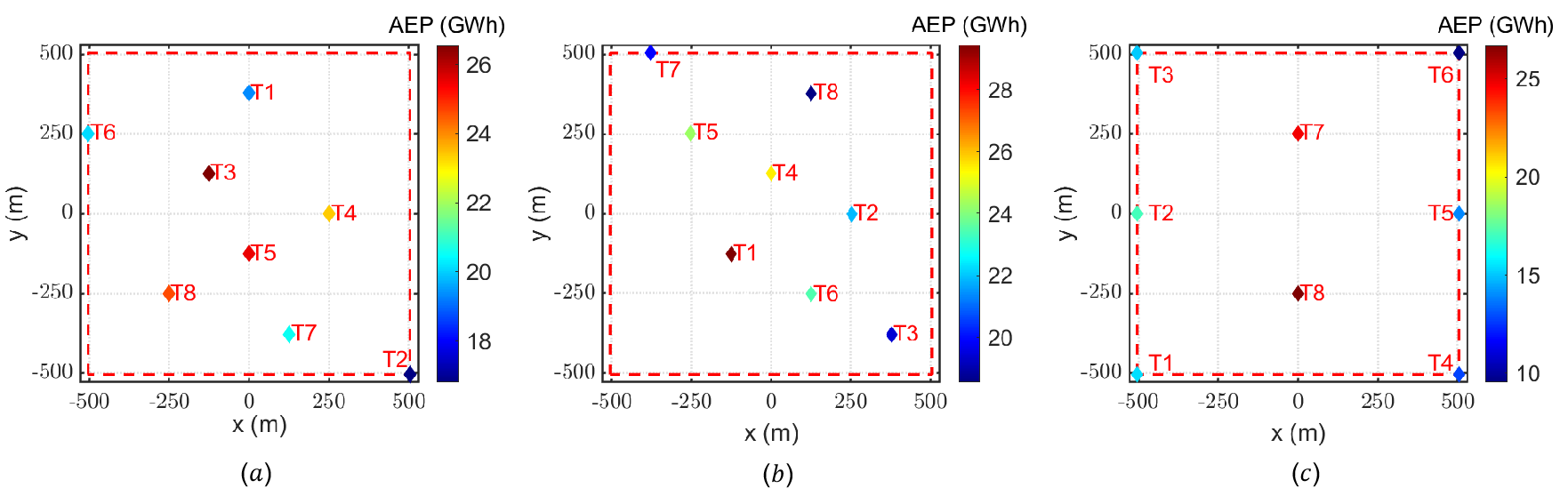}
    \caption{Case III: AEP distribution for layouts obtained from $(a)$ SBO $(b)$ CFD-GA and $(c)$ baseline.}
    \label{fig:case3layout}
\end{figure}

\begin{table}
    \centering
    \caption{\label{tab:results}AEP and CFD calls of optimized layouts using different optimization strategy for 3 cases}
    \begin{tabular}{c c c c c c c}
         \hline
           & \multicolumn{2}{c}{Case I}& \multicolumn{2}{c}{Case II}& \multicolumn{2}{c}{Case III}  \\
         \cline{2-7}
           &CFD calls &AEP (GWh) &CFD calls &AEP (GWh) &CFD calls &AEP (GWh) \\
         \hline
           SBO       &437 &202.740 &400 &120.374 &399 &178.037 \\
           CFD-GA    &961 &204.400 &793 &126.707 &885 &183.054 \\
           Baseline  &- &141.324 &- &97.287 &- &135.732 \\
         \hline
    \end{tabular}
    \label{tab:results}
\end{table}

\begin{figure}
    \centering
    \includegraphics[width=1.0\textwidth]{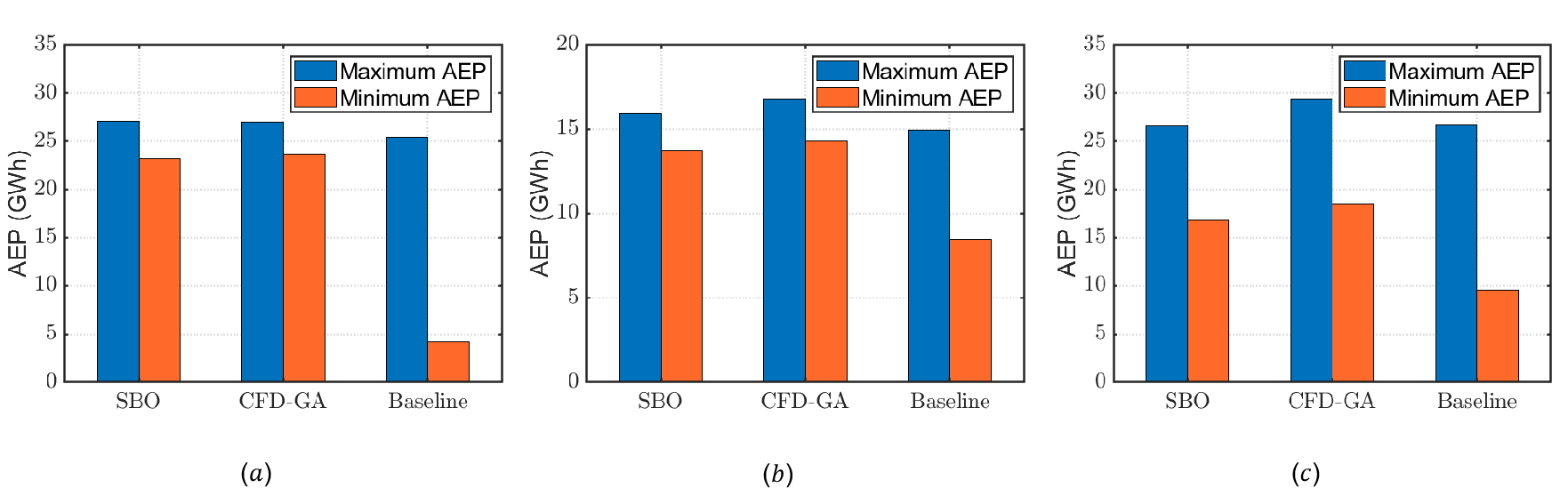}
    \caption{Comparison of maximum and minimum AEP of the turbines in the wind farm for optimal layouts for $(a)$ Case I, $(b)$ Case II, and $(c)$ Case III.}
    \label{fig:casedifference}
\end{figure}

\begin{figure*}
    \centering
    \includegraphics[width=0.5\textwidth]{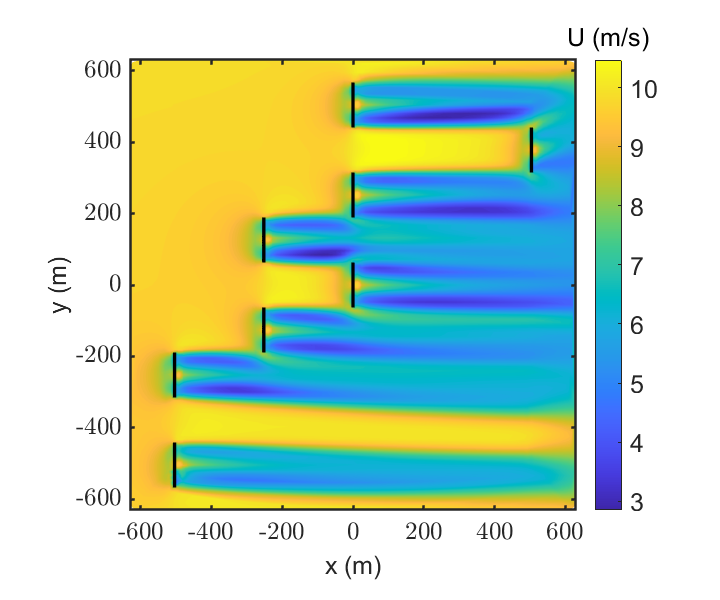}
    \caption{Velocity magnitude at rotor hub height of optimized layout for case I}
    \label{fig:case1velocity}
\end{figure*}

\begin{figure*}
    \centering
    \includegraphics[width=1.0\textwidth]{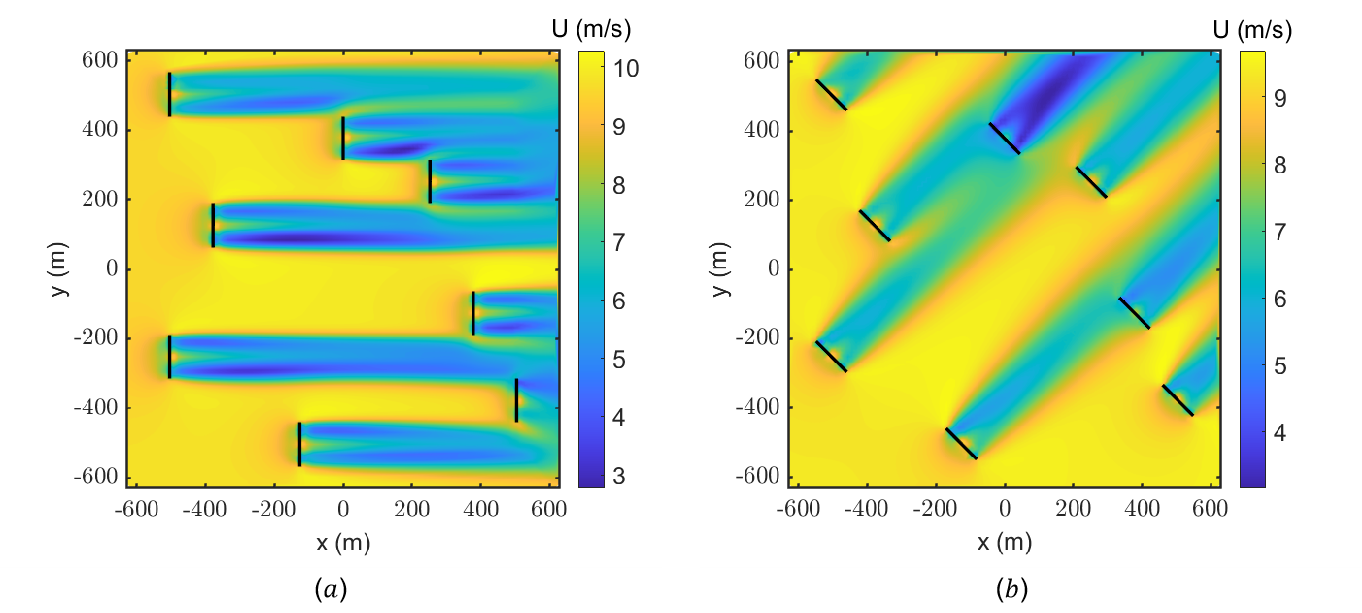}
    \caption{Velocity magnitude at rotor hub height of optimized layout for case II. $(a)$ West, and $(b)$ South-West. }
    \label{fig:case2velocity}
\end{figure*}

\begin{figure*}
    \centering
    \includegraphics[width=1.0\textwidth]{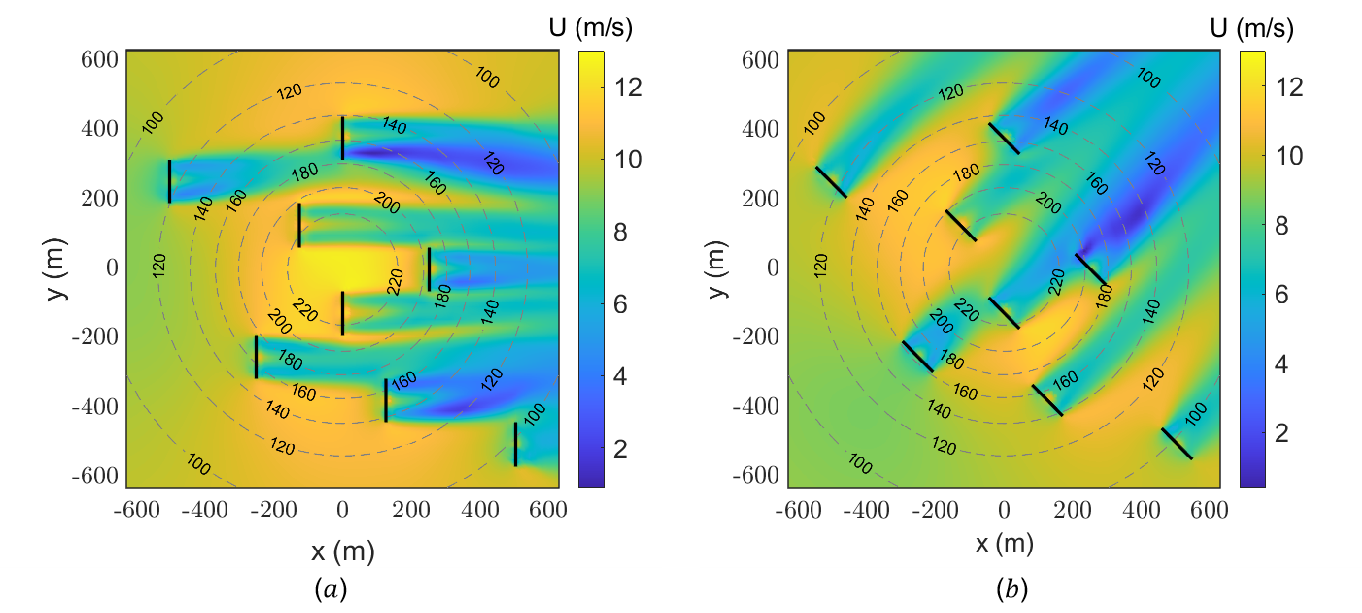}
    \caption{Velocity magnitude at rotor hub height of optimized layout for case III. $(a)$ West, and $(b)$ South-West. The gray dashed lines indicate the iso-height lines of the terrain.}
    \label{fig:case3velocity}
\end{figure*}

The optimized and baseline layouts with their AEP distribution are shown in figure \ref{fig:case1layout} to \ref{fig:case3layout}, and the number of CFD calls for different optimization strategies along with their AEPs are listed in table \ref{tab:results}. 
Compared to baseline layouts, optimized layouts obtained from this framework and CFD-GA method both achieve significant AEP improvement. 
Due to the multi-modality of the WFLO problem, the optimization process can get stuck in a local minimum, thus the optimal layout results obtained from the CFD-GA and SBO strategies are usually not identical even for the same case.
Nevertheless, the AEP of optimized layouts obtained from SBO are almost the same from that obtained by CFD-GA. 
In terms of computational cost, the numbers of CFD calls for SBO are $437$, $400$ and $399$, compared to $961$, $793$ and $885$ for CFD-GA in case I, case II and case III, respectively. 
These CFD simulations are executed on two desktop computers, both equipped with AMD EPYC 7532 processor with 32 cores. 
For the three wind farm cases, it takes approximately 72 h, 336 h and 360 h to obtain optimized layouts using CFD-GA method while it only takes approximately 36 h, 172 h and 190 h using the developed framework.
Above all, we see that the surrogate-based optimization framework takes only about half computational cost of the CFD-based method to find optimal designs which produce almost the same AEP.

We also compare the maximum and minimum AEP of the turbines in the wind farm for different layouts in figure \ref{fig:casedifference}.
It is argued that, in a well-designed wind farm layout, the AEP among the different turbines should be distributed as even as possible, such that each turbine contribute a fair share to the total AEP.
As shown in figure \ref{fig:casedifference}, the difference between maximum and minimum AEP in optimized layouts (obtained from both SBO and CFD-GA) is much smaller than that in baseline layouts for the two cases, indicating that AEP of optimized layouts becomes much more evenly-distributed. 
For case III, the difference between the maximum and minimum AEP is significantly larger. 
From figure \ref{fig:case3layout}, it is observed that the turbine associated with maximum AEP is located near the hill top, where the wind velocity increases due to the terrain effects. 
Nevertheless, the optimized layouts still feature more even AEP distribution than the reference layout.

We further show the velocity fields at the hub height of optimized layouts obtained from SBO in figures \ref{fig:case1velocity}-\ref{fig:case3velocity}. 
For case I where only a single wind direction is considered, we notice that in SBO's optimized layout, all turbines are placed away from other turbines' wakes.
In case II, the western inflow with wind speed of 11.3 m/s and occurrence frequency of 0.21 is the most dominant wind direction, and the southwestern inflow with wind speed of 10.9 m/s and occurrence frequency of 0.203 is the secondary dominant wind direction, as shown in the wind rose in figure \ref{fig:windrose}. 
It is shown in figure \ref{fig:case2layout} that most turbines are located away from other turbines' wake zone along the top two dominant wind directions. 
In case III, the effects of the terrain is considered.
With the Gaussian hill at the center of the computational domain, the wind turbines in the optimized layouts are clustered toward the center to take advantage of the higher wind velocity.
The ability of taking the terrain effects into consideration equips the current CFD-based WFLO method with more flexibility compared to those based on analytical wake models.

\section{Conclusions}
\label{sec:conclusion}
We have developed a surrogate-based framework for wind farm layout optimization with higher fidelity wake modeling methods, with the objective of maximizing the annual energy production. 
The surrogate is built by the Kriging model, which is trained using CFD simulations by treating the turbines as actuator disks. 
The surrogate model is then integrated with the genetic algorithm as the optimizer.
During the optimization process, the wind farm layout dataset is updated adaptively with the intermediate design during each iteration added into it until the algorithm converges to the optimal layout.
To assess its performance, we have tested the proposed SBO framework in three WFLO cases with different wind distributions and terrain.
It is observed that the optimized layouts obtained from this self-adaptive framework generates almost similar AEP with CFD-GA direct optimization (without surrogate), and significantly outperforms the conventional staggered layout along the major wind direction. 
Aided by the surrogate modeling technique, the computational time of SBO framework is only half of that for the direct method, thus allowing the CFD-based WFLO to be carried out with manageable cost.

While the current study has demonstrated the feasibility of using Kriging model to accelerate high-fidelity wind farm layout optimization, there are still some limitations in the framework that need further improvement.
First, the use of Gaussian process for building surrogate models is known to be ``cursed" for high-dimensional data \citep{binois2022survey}. Innovative techniques for sovling the scalability problem are in urgent need if this SBO framework is to be applied in WFLO tasks with larger number of wind turbines.
Second, current work considers a single objective of maximizing AEP. Additional objectives, such as minimizing fatigue loads, minimizing construction cost, etc, can be incorporated in future studies.
In addition, multi-fidelity methods have shown great potential in a variety of complex design problems \citep{forrester2007multi,kou2019multi,jasa2022effectively}. For WFLO, the results from low-fidelity analytical wake models can also be expected to play an important role in further improving the optimization efficiency using the multi-fidelity optimization framework.

\section*{Acknowledgments}
DZ, ZH, KZ acknowledge financial support from the Innovation Program of Shanghai Municipal Education Commission (no. 2019-01-07-00-02-E00066), National Science Foundation of China (grant numbers: 12202271, 52122110, 42076210), Program for Intergovernmental International S\&T Cooperation Projects of Shanghai Municipality, China (grant no. 22160710200), and the Oceanic Interdisciplinary Program of Shanghai Jiao Tong University (grant no. SL2020PT201).

\bibliographystyle{unsrtnat}
\bibliography{references}  

\begin{thebibliography}{56}
\providecommand{\natexlab}[1]{#1}
\providecommand{\url}[1]{\texttt{#1}}
\expandafter\ifx\csname urlstyle\endcsname\relax
  \providecommand{\doi}[1]{doi: #1}\else
  \providecommand{\doi}{doi: \begingroup \urlstyle{rm}\Url}\fi

\bibitem[Barthelmie et~al.(2009)Barthelmie, Hansen, Frandsen, Rathmann,
  Schepers, Schlez, Phillips, Rados, Zervos, Politis,
  et~al.]{barthelmie2009modelling}
Rebecca~Jane Barthelmie, K~Hansen, Sten~Tron{\ae}s Frandsen, Ole Rathmann,
  JG~Schepers, W~Schlez, J~Phillips, K~Rados, A~Zervos, ESa Politis, et~al.
\newblock Modelling and measuring flow and wind turbine wakes in large wind
  farms offshore.
\newblock \emph{Wind Energy: An International Journal for Progress and
  Applications in Wind Power Conversion Technology}, 12\penalty0 (5):\penalty0
  431--444, 2009.

\bibitem[Reddy(2020)]{reddy2020wind}
Sohail~R Reddy.
\newblock Wind farm layout optimization ({WindFLO}): An advanced framework for
  fast wind farm analysis and optimization.
\newblock \emph{Applied Energy}, 269:\penalty0 115090, 2020.

\bibitem[Dong et~al.(2021)Dong, Zhang, and Zhao]{dong2021intelligent}
Hongyang Dong, Jincheng Zhang, and Xiaowei Zhao.
\newblock Intelligent wind farm control via deep reinforcement learning and
  high-fidelity simulations.
\newblock \emph{Applied Energy}, 292:\penalty0 116928, 2021.

\bibitem[Thomas et~al.(2022{\natexlab{a}})Thomas, Baker, Malisani, Quaeghebeur,
  Perez-Moreno, Jasa, Bay, Tilli, Bieniek, Robinson,
  et~al.]{thomas2022comparison}
Jared~J Thomas, Nicholas~F Baker, Paul Malisani, Erik Quaeghebeur,
  Sebastian~Sanchez Perez-Moreno, John Jasa, Christopher Bay, Federico Tilli,
  David Bieniek, Nick Robinson, et~al.
\newblock A comparison of eight optimization methods applied to a wind farm
  layout optimization problem.
\newblock \emph{Wind Energy Science}, 8:\penalty0 865–891,
  2022{\natexlab{a}}.

\bibitem[Mosetti et~al.(1994)Mosetti, Poloni, and
  Diviacco]{mosetti1994optimization}
GPCDB Mosetti, Carlo Poloni, and Bruno Diviacco.
\newblock Optimization of wind turbine positioning in large windfarms by means
  of a genetic algorithm.
\newblock \emph{Journal of Wind Engineering and Industrial Aerodynamics},
  51\penalty0 (1):\penalty0 105--116, 1994.

\bibitem[Ozturk and Norman(2004)]{ozturk2004heuristic}
U~Aytun Ozturk and Bryan~A Norman.
\newblock Heuristic methods for wind energy conversion system positioning.
\newblock \emph{Electric Power Systems Research}, 70\penalty0 (3):\penalty0
  179--185, 2004.

\bibitem[Grady et~al.(2005)Grady, Hussaini, and Abdullah]{grady2005placement}
SA~Grady, MY~Hussaini, and Makola~M Abdullah.
\newblock Placement of wind turbines using genetic algorithms.
\newblock \emph{Renewable Energy}, 30\penalty0 (2):\penalty0 259--270, 2005.

\bibitem[Mora et~al.(2007)Mora, Baron, Santos, and Payan]{mora2007evolutive}
Jose~Castro Mora, Jose M~Calero Baron, Jesus M~Riquelme Santos, and
  Manuel~Burgos Payan.
\newblock An evolutive algorithm for wind farm optimal design.
\newblock \emph{Neurocomputing}, 70\penalty0 (16-18):\penalty0 2651--2658,
  2007.

\bibitem[Zhang et~al.(2011)Zhang, Hou, and Wang]{zhang2011fast}
Changshui Zhang, Guangdong Hou, and Jun Wang.
\newblock A fast algorithm based on the submodular property for optimization of
  wind turbine positioning.
\newblock \emph{Renewable Energy}, 36\penalty0 (11):\penalty0 2951--2958, 2011.

\bibitem[Ero{\u{g}}lu and Se{\c{c}}kiner(2012)]{erouglu2012design}
Yunus Ero{\u{g}}lu and Serap~Ulusam Se{\c{c}}kiner.
\newblock Design of wind farm layout using ant colony algorithm.
\newblock \emph{Renewable Energy}, 44:\penalty0 53--62, 2012.

\bibitem[Chen et~al.(2013)Chen, Li, Jin, and Song]{chen2013wind}
Ying Chen, Hua Li, Kai Jin, and Qing Song.
\newblock Wind farm layout optimization using genetic algorithm with different
  hub height wind turbines.
\newblock \emph{Energy Conversion and Management}, 70:\penalty0 56--65, 2013.

\bibitem[Park and Law(2015)]{park2015layout}
Jinkyoo Park and Kincho~H Law.
\newblock Layout optimization for maximizing wind farm power production using
  sequential convex programming.
\newblock \emph{Applied Energy}, 151:\penalty0 320--334, 2015.

\bibitem[Shakoor et~al.(2015)Shakoor, Hassan, Raheem, and
  Rasheed]{shakoor2015modelling}
Rabia Shakoor, Mohammad~Yusri Hassan, Abdur Raheem, and Nadia Rasheed.
\newblock The modelling of wind farm layout optimization for the reduction of
  wake losses.
\newblock \emph{Indian Journal of Science and Technology}, 8\penalty0
  (17):\penalty0 1--9, 2015.

\bibitem[Hou et~al.(2016)Hou, Hu, Chen, Soltani, and Chen]{hou2016optimization}
Peng Hou, Weihao Hu, Cong Chen, Mohsen Soltani, and Zhe Chen.
\newblock Optimization of offshore wind farm layout in restricted zones.
\newblock \emph{Energy}, 113:\penalty0 487--496, 2016.

\bibitem[Gebraad et~al.(2017)Gebraad, Thomas, Ning, Fleming, and
  Dykes]{gebraad2017maximization}
Pieter Gebraad, Jared~J Thomas, Andrew Ning, Paul Fleming, and Katherine Dykes.
\newblock Maximization of the annual energy production of wind power plants by
  optimization of layout and yaw-based wake control.
\newblock \emph{Wind Energy}, 20\penalty0 (1):\penalty0 97--107, 2017.

\bibitem[Parada et~al.(2017)Parada, Herrera, Flores, and
  Parada]{parada2017wind}
Leandro Parada, Carlos Herrera, Paulo Flores, and Victor Parada.
\newblock Wind farm layout optimization using a gaussian-based wake model.
\newblock \emph{Renewable Energy}, 107:\penalty0 531--541, 2017.

\bibitem[Pillai et~al.(2017)Pillai, Chick, Khorasanchi, Barbouchi, and
  Johanning]{pillai2017application}
Ajit~C Pillai, John Chick, Mahdi Khorasanchi, Sami Barbouchi, and Lars
  Johanning.
\newblock Application of an offshore wind farm layout optimization methodology
  at middelgrunden wind farm.
\newblock \emph{Ocean Engineering}, 139:\penalty0 287--297, 2017.

\bibitem[Kirchner-Bossi and Port{\'e}-Agel(2018)]{kirchner2018realistic}
Nicolas Kirchner-Bossi and Fernando Port{\'e}-Agel.
\newblock Realistic wind farm layout optimization through genetic algorithms
  using a {Gaussian} wake model.
\newblock \emph{Energies}, 11\penalty0 (12):\penalty0 3268, 2018.

\bibitem[Stanley and Ning(2019)]{stanley2019massive}
Andrew~PJ Stanley and Andrew Ning.
\newblock Massive simplification of the wind farm layout optimization problem.
\newblock \emph{Wind Energy Science}, 4\penalty0 (4):\penalty0 663--676, 2019.

\bibitem[Quan and Kim(2019)]{quan2019greedy}
Ning Quan and Harrison~M Kim.
\newblock Greedy robust wind farm layout optimization with feasibility
  guarantee.
\newblock \emph{Engineering Optimization}, 51\penalty0 (7):\penalty0
  1152--1167, 2019.

\bibitem[Cruz and Carmo(2020)]{cruz2020wind}
Lu{\'\i}s Eduardo~Boni Cruz and Bruno~Souza Carmo.
\newblock Wind farm layout optimization based on {CFD} simulations.
\newblock \emph{Journal of the Brazilian Society of Mechanical Sciences and
  Engineering}, 42:\penalty0 1--18, 2020.

\bibitem[Antonini et~al.(2020)Antonini, Romero, and Amon]{antonini2020optimal}
Enrico~GA Antonini, David~A Romero, and Cristina~H Amon.
\newblock Optimal design of wind farms in complex terrains using computational
  fluid dynamics and adjoint methods.
\newblock \emph{Applied Energy}, 261:\penalty0 114426, 2020.

\bibitem[Stanley et~al.(2020)Stanley, King, and Ning]{stanley2020wind}
Andrew~PJ Stanley, Jennifer King, and Andrew Ning.
\newblock Wind farm layout optimization with loads considerations.
\newblock In \emph{Journal of Physics: Conference Series}, volume 1452, page
  012072. IOP Publishing, 2020.

\bibitem[Gagakuma et~al.(2021)Gagakuma, Stanley, and
  Ning]{gagakuma2021reducing}
Bertelsen Gagakuma, Andrew~PJ Stanley, and Andrew Ning.
\newblock Reducing wind farm power variance from wind direction using wind farm
  layout optimization.
\newblock \emph{Wind Engineering}, 45\penalty0 (6):\penalty0 1517--1530, 2021.

\bibitem[Liu et~al.(2021)Liu, Fan, Wang, and Peng]{liu2021genetic}
Zhenqing Liu, Shuanglong Fan, Yize Wang, and Jie Peng.
\newblock Genetic-algorithm-based layout optimization of an offshore wind farm
  under real seabed terrain encountering an engineering cost model.
\newblock \emph{Energy Conversion and Management}, 245:\penalty0 114610, 2021.

\bibitem[Thomas et~al.(2022{\natexlab{b}})Thomas, McOmber, and
  Ning]{thomas2022wake}
Jared~J Thomas, Spencer McOmber, and Andrew Ning.
\newblock Wake expansion continuation: Multi-modality reduction in the wind
  farm layout optimization problem.
\newblock \emph{Wind Energy}, 25\penalty0 (4):\penalty0 678--699,
  2022{\natexlab{b}}.

\bibitem[Jensen(1983)]{jensen1983note}
Niels~Otto Jensen.
\newblock \emph{A note on wind generator interaction}.
\newblock Report Risø-M-2411, Risø National Laboratory, Roskilde, Denmark,
  1983.

\bibitem[Larsen(1988)]{larsen1988simple}
Gunner~Chr Larsen.
\newblock \emph{A simple wake calculation procedure}.
\newblock Report Riso-M-2760, Ris{\o} National Laboratory, Roskilde, Denmark,
  1988.

\bibitem[Frandsen et~al.(2006)Frandsen, Barthelmie, Pryor, Rathmann, Larsen,
  H{\o}jstrup, and Th{\o}gersen]{frandsen2006analytical}
Sten Frandsen, Rebecca Barthelmie, Sara Pryor, Ole Rathmann, S{\o}ren Larsen,
  J{\o}rgen H{\o}jstrup, and Morten Th{\o}gersen.
\newblock Analytical modelling of wind speed deficit in large offshore wind
  farms.
\newblock \emph{Wind Energy: An International Journal for Progress and
  Applications in Wind Power Conversion Technology}, 9\penalty0 (1-2):\penalty0
  39--53, 2006.

\bibitem[Mittal et~al.(2016)Mittal, Sreenivas, Taylor, Hereth, and
  Hilbert]{mittal2016blade}
Anshul Mittal, Kidambi Sreenivas, Lafayette~K Taylor, Levi Hereth, and
  Christopher~B Hilbert.
\newblock Blade-resolved simulations of a model wind turbine: effect of
  temporal convergence.
\newblock \emph{Wind Energy}, 19\penalty0 (10):\penalty0 1761--1783, 2016.

\bibitem[Liu et~al.(2017)Liu, Xiao, Incecik, Peyrard, and
  Wan]{liu2017establishing}
Yuanchuan Liu, Qing Xiao, Atilla Incecik, Christophe Peyrard, and Decheng Wan.
\newblock Establishing a fully coupled {CFD} analysis tool for floating
  offshore wind turbines.
\newblock \emph{Renewable Energy}, 112:\penalty0 280--301, 2017.

\bibitem[de~Oliveira et~al.(2022)de~Oliveira, Puraca, and Carmo]{de2022blade}
Marielle de~Oliveira, Rodolfo~Curci Puraca, and Bruno~Souza Carmo.
\newblock Blade-resolved numerical simulations of the nrel offshore 5 mw
  baseline wind turbine in full scale: A study of proper solver configuration
  and discretization strategies.
\newblock \emph{Energy}, 254:\penalty0 124368, 2022.

\bibitem[Zhang et~al.(2023)Zhang, Kuang, Han, Zhou, Zhao, Bao, Duan, Tu, Chen,
  and Chen]{zhang2023comparative}
Zhihao Zhang, Limin Kuang, Zhaolong Han, Dai Zhou, Yongsheng Zhao, Yan Bao, Lei
  Duan, Jiahuang Tu, Yaoran Chen, and Mingsheng Chen.
\newblock Comparative analysis of bent and basic winglets on performance
  improvement of horizontal axis wind turbines.
\newblock \emph{Energy}, 281:\penalty0 128252, 2023.

\bibitem[Stevens and Meneveau(2017)]{stevens2017flow}
Richard J. A.~M. Stevens and Charles Meneveau.
\newblock Flow structure and turbulence in wind farms.
\newblock \emph{Annual Review of Fluid Mechanics}, 49:\penalty0 311--339, 2017.

\bibitem[Stevens et~al.(2018)Stevens, Mart{\'\i}nez-Tossas, and
  Meneveau]{stevens2018comparison}
Richard J. A.~M. Stevens, Luis~A Mart{\'\i}nez-Tossas, and Charles Meneveau.
\newblock Comparison of wind farm large eddy simulations using actuator disk
  and actuator line models with wind tunnel experiments.
\newblock \emph{Renewable Energy}, 116:\penalty0 470--478, 2018.

\bibitem[Shapiro et~al.(2022)Shapiro, Starke, and Gayme]{shapiro2022turbulence}
Carl~R Shapiro, Genevieve~M Starke, and Dennice~F Gayme.
\newblock Turbulence and control of wind farms.
\newblock \emph{Annual Review of Control, Robotics, and Autonomous Systems},
  5:\penalty0 579--602, 2022.

\bibitem[Sanderse et~al.(2011)Sanderse, Van~der Pijl, and
  Koren]{sanderse2011review}
Benjamin Sanderse, SP~Van~der Pijl, and Barry Koren.
\newblock Review of computational fluid dynamics for wind turbine wake
  aerodynamics.
\newblock \emph{Wind Energy}, 14\penalty0 (7):\penalty0 799--819, 2011.

\bibitem[Svenning(2010)]{svenning2010implementation}
Erik Svenning.
\newblock Implementation of an actuator disk in openfoam.
\newblock \emph{Chalmers University of Technology}, 2010.

\bibitem[Mart{\'\i}nez-Tossas et~al.(2015)Mart{\'\i}nez-Tossas, Churchfield,
  and Leonardi]{martinez2015large}
Luis~A Mart{\'\i}nez-Tossas, Matthew~J Churchfield, and Stefano Leonardi.
\newblock Large eddy simulations of the flow past wind turbines: actuator line
  and disk modeling.
\newblock \emph{Wind Energy}, 18\penalty0 (6):\penalty0 1047--1060, 2015.

\bibitem[Krige(1951)]{krige1951statistical}
Daniel~G Krige.
\newblock \emph{A statistical approach to some mine valuation and allied
  problems on the Witwatersrand}.
\newblock PhD thesis, University of the Witwatersrand, 1951.

\bibitem[Karl(2010)]{karl2010spatial}
Jason~W Karl.
\newblock Spatial predictions of cover attributes of rangeland ecosystems using
  regression kriging and remote sensing.
\newblock \emph{Rangeland Ecology \& Management}, 63\penalty0 (3):\penalty0
  335--349, 2010.

\bibitem[Casella(2020)]{casella2020surrogate}
S~G Casella.
\newblock Surrogate based optimisation for hull shapes with dakota toolkit.
\newblock Master's thesis, Delft University of Technology, 2020.

\bibitem[Montoya et~al.(2022)Montoya, Hern{\'a}ndez, and
  Kareem]{montoya2022aero}
M~Cid Montoya, S~Hern{\'a}ndez, and Ahsan Kareem.
\newblock Aero-structural optimization-based tailoring of bridge deck geometry.
\newblock \emph{Engineering Structures}, 270:\penalty0 114067, 2022.

\bibitem[Bai et~al.(2022)Bai, Ju, Wang, Zhou, and Liu]{bai2022wind}
Fangyun Bai, Xinglong Ju, Shouyi Wang, Wenyong Zhou, and Feng Liu.
\newblock Wind farm layout optimization using adaptive evolutionary algorithm
  with {Monte Carlo Tree Search} reinforcement learning.
\newblock \emph{Energy Conversion and Management}, 252:\penalty0 115047, 2022.

\bibitem[McKay et~al.(2000)McKay, Beckman, and Conover]{mckay2000comparison}
Michael~D McKay, Richard~J Beckman, and William~J Conover.
\newblock A comparison of three methods for selecting values of input variables
  in the analysis of output from a computer code.
\newblock \emph{Technometrics}, 42\penalty0 (1):\penalty0 55--61, 2000.

\bibitem[Richmond et~al.(2019)Richmond, Antoniadis, Wang, Kolios, Al-Sanad, and
  Parol]{richmond2019evaluation}
Mark Richmond, A~Antoniadis, Lin Wang, Athanasios Kolios, S~Al-Sanad, and
  Jafarali Parol.
\newblock Evaluation of an offshore wind farm computational fluid dynamics
  model against operational site data.
\newblock \emph{Ocean Engineering}, 193:\penalty0 106579, 2019.

\bibitem[Burton et~al.(2011)Burton, Jenkins, Sharpe, and
  Bossanyi]{burton2011wind}
Tony Burton, Nick Jenkins, David Sharpe, and Ervin Bossanyi.
\newblock \emph{Wind energy handbook}.
\newblock John Wiley \& Sons, 2011.

\bibitem[Goldstein(1929)]{goldstein1929vortex}
Sydney Goldstein.
\newblock On the vortex theory of screw propellers.
\newblock \emph{Proceedings of the Royal Society of London. Series A,
  Containing Papers of a Mathematical and Physical Character}, 123\penalty0
  (792):\penalty0 440--465, 1929.

\bibitem[Dalbey et~al.(2022)Dalbey, Eldred, Geraci, Jakeman, Maupin, Monschke,
  Seidl, Tran, Menhorn, and Zeng]{dalbey2022dakota}
Keith Dalbey, Michael Eldred, Gianluca Geraci, John Jakeman, Kathryn Maupin,
  Jason~A Monschke, Daniel Seidl, Anh Tran, Friedrich Menhorn, and Xiaoshu
  Zeng.
\newblock Dakota, a multilevel parallel object-oriented framework for design
  optimization, parameter estimation, uncertainty quantification, and
  sensitivity analysis: Version 6.16 theory manual.
\newblock Technical report, Sandia National Lab.(SNL-NM), Albuquerque, NM
  (United States), 2022.

\bibitem[Adams et~al.(2020)Adams, Bohnhoff, Dalbey, Ebeida, Eddy, Eldred,
  Hooper, Hough, Hu, Jakeman, et~al.]{adams2020dakota}
Brian~M Adams, William~J Bohnhoff, Keith~R Dalbey, Mohamed~S Ebeida, John~P
  Eddy, Michael~S Eldred, Russell~W Hooper, Patricia~D Hough, Kenneth~T Hu,
  John~D Jakeman, et~al.
\newblock Dakota, a multilevel parallel object-oriented framework for design
  optimization, parameter estimation, uncertainty quantification, and
  sensitivity analysis: version 6.13 user's manual.
\newblock Technical report, Sandia National Lab.(SNL-NM), Albuquerque, NM
  (United States), 2020.

\bibitem[Thelen(2016)]{thelen2016surrogate}
Andrew Thelen.
\newblock \emph{Surrogate-based design optimization of dual-rotor wind turbines
  using steady RANS equations}.
\newblock PhD thesis, Iowa State University, 2016.

\bibitem[Jonkman et~al.(2009)Jonkman, Butterfield, Musial, and
  Scott]{jonkman2009definition}
Jason Jonkman, Sandy Butterfield, Walter Musial, and George Scott.
\newblock {Definition of a 5-MW reference wind turbine for offshore system
  development}.
\newblock Technical report, National Renewable Energy Lab.(NREL), Golden, CO
  (United States), 2009.

\bibitem[Binois and Wycoff(2022)]{binois2022survey}
Mickael Binois and Nathan Wycoff.
\newblock A survey on high-dimensional gaussian process modeling with
  application to bayesian optimization.
\newblock \emph{ACM Transactions on Evolutionary Learning and Optimization},
  2\penalty0 (2):\penalty0 1--26, 2022.

\bibitem[Forrester et~al.(2007)Forrester, S{\'o}bester, and
  Keane]{forrester2007multi}
Alexander~IJ Forrester, Andr{\'a}s S{\'o}bester, and Andy~J Keane.
\newblock Multi-fidelity optimization via surrogate modelling.
\newblock \emph{Proceedings of the Royal Society A: Mathematical, Physical and
  Engineering Sciences}, 463\penalty0 (2088):\penalty0 3251--3269, 2007.

\bibitem[Kou and Zhang(2019)]{kou2019multi}
Jiaqing Kou and Weiwei Zhang.
\newblock Multi-fidelity modeling framework for nonlinear unsteady aerodynamics
  of airfoils.
\newblock \emph{Applied Mathematical Modelling}, 76:\penalty0 832--855, 2019.

\bibitem[Jasa et~al.(2022)Jasa, Bortolotti, Zalkind, and
  Barter]{jasa2022effectively}
John Jasa, Pietro Bortolotti, Daniel Zalkind, and Garrett Barter.
\newblock Effectively using multifidelity optimization for wind turbine design.
\newblock \emph{Wind Energy Science}, 7\penalty0 (3):\penalty0 991--1006, 2022.

\end{thebibliography}






\end{document}